\documentclass[aps,prl,twocolumn,footinbib,longbibliography,superscriptaddress]{revtex4-1}
\usepackage{amsmath,amssymb}
\usepackage{bm}
\usepackage{epsfig}
\usepackage{natbib}
\usepackage{hyperref}
\usepackage{color}
\usepackage{graphicx}

 \hypersetup{pdfstartview={FitH},pdfpagemode={UseNone},
            colorlinks,linkcolor=red, citecolor=blue, urlcolor=blue,
            bookmarks=true, bookmarksopen=true, pdfnewwindow=true}

\newcommand{\rmi}{{\rm i}}

\newcommand{\e}{{\rm e}}

\begin{document}

\title{Multiple quantum wells for $\mathcal{PT}$-symmetric phononic crystals}

\author{\firstname{A.~V.} \surname{Poshakinskiy}}
\email{poshakinskiy@mail.ioffe.ru}
\affiliation{Ioffe  Institute, St.~Petersburg 194021, Russia}
\author{\firstname{A.~N.} \surname{Poddubny}}
\affiliation{Ioffe  Institute, St.~Petersburg 194021, Russia}
\author{\firstname{A.} \surname{Fainstein}}
\affiliation{Centro Atomico Bariloche and Instituto Balseiro, C.N.E.A., 8400 S. C. de Bariloche, R. N., Argentina}

\begin{abstract}

We demonstrate that the parity-time symmetry for sound is realized in the laser-pumped multiple-quantum-well structure. Breaking of the parity-time symmetry for the phonons with wave vectors corresponding to the Bragg condition makes the structure a highly-selective acoustic wave amplifier. 
Single-mode distributed feedback phonon lasing is predicted for structures with realistic parameters.

\end{abstract}
 \maketitle

{\it Introduction.}---Parity-time ($\mathcal{PT}$) symmetry, initially proposed as a concept in quantum mechanics~\cite{Bender1998}, is now being extensively studied in optics~\cite{Zyablovsky2014ufn,Suchkov2015,Konotop2016}. A $\mathcal{PT}$-symmetric optical system is the one with a dielectric permittivity distribution that is invariant under simultaneous operations of spatial inversion and complex conjugation. 
 In such structure, any spatial region with a loss is mirrored by a region of gain. 
Therefore, the processes of light absorption and amplification can be compensated, and the frequencies of eigen optical modes can be real. 
An elementary $\mathcal{PT}$-symmetric structure is a pair of loss and gain cavities. It allows one to realize power oscillations~\cite{Ruter2010} and tailor the nonlinear response~\cite{Hodaei2014,Peng2014b}. 
In $\mathcal{PT}$-symmetric photonic crystals Bragg amplification boosts
unidirectional invisibility~\cite{Lin2011,Feng2012,Regensburger2012}, coherent perfect absorption \cite{Stone2011} and  single-mode lasing~\cite{Feng2014}.

Recently, the concept of $\mathcal{PT}$ symmetry has been introduced into acoustics~\cite{Zhu2014}. While the general ideas directly mimic optical structures,  controllable sound amplification/attenuation requires completely different physical mechanisms than those for light. So far, the experimental demonstration of $\mathcal{PT}$ symmetry in mechanical system has been limited to 
macroscopic pendula driven by electromagnets~\cite{Bender2013}. Another macroscopic structure based on bulk semiconductor acoustic Cherenkov amplifiers been suggested in Ref.~\cite{Lu2016}.  
A promising mechanism of sound amplification and attenuation at the nanoscale is the optomechanical cooling and heating~\cite{Kippenberg2014}. So far this approach has been exploited just for two coupled mechanical nanoresonators~\cite{Xu2015b}.  Here, we show that semiconductor multiple quantum well (MQW) structures enable realization of   $\mathcal{PT}$-symmetric phononic crystals, that, similarly to their optical counterparts,  allow for new possibilities of   sound manipulation.

The quantum well structures form an established platform successfully used for quantum cascade lasers~\cite{Capasso1994}, sasers~\cite{Kent2006,Maryam2013,Beardsley2010},  phononic crystals and nanoresonators~\cite{Kimura2007,Fainstein2013}.  Strong enhancement of photoelastic light-sound interaction at the exciton resonance has recently been demonstrated~\cite{Jusserand2015}. Here,  we predict that  a laser-pumped QW with realistic parameters acts as a phonon amplifier/attenuator depending on the detuning of the pump laser from an exciton resonance. Combining QWs that have exciton frequency positively and negatively detuned from the laser frequency in the same structure we can realize $\mathcal{PT}$ symmetry for sound.

{\it Sound transmission and reflection from a single QW.}---We begin with a demonstration that a single QW pumped by a laser light of the frequency close to the exciton resonance can be used for controllable phonon amplification and attenuation. The Hamiltonian describing an interaction of the longitudinal acoustic phonons with excitons in a QW reads  
\begin{equation}\label{eq:H}
 H = \omega_x b^\dag b + \sum_k \Omega_k^{\vphantom{\dag}} a_k^\dag a_k^{\vphantom{\dag}} + \sum_k b^\dag b (g_k a_k + g_k^* a_k^\dag) \,,
\end{equation}
where $\omega_x$ is the QW exciton resonance frequency, $\Omega_k = s|k|$ is the phonon dispersion with $s$ being the speed of sound, $a_k$ and $b$ are the annihilation operators for the phonons and excitons, respectively, and $\hbar =1$. The last term of Eq.~\eqref{eq:H} stands for the exciton-phonon interaction due to the deformation potential mechanism~\cite{Cardona}. The matrix element of the interaction reads $g_k = \rmi k\Xi_k /\sqrt{2\rho\Omega_k S}$, where $\Xi_k = \int (\Xi_e |\psi_e(z)|^2+ \Xi_h |\psi_h(z)|^2)\,\e^{-\rmi k z} {\rm d}z$ with $\psi_{e(h)}$ and $\Xi_{e(h)}$ being the electron (hole) confinement wave functions and deformation potential constants, respectively, $\rho$ is the mass density, and $S$ is the  sample (normalization) area~\cite{Jusserand2013,Poddubny2014}.

When a QW is excited by a laser of the frequency $\omega_L$ close to the exciton resonance, the coherent exciton population is created. This can be taken into account by making the substitution $b \to b_L + b$ in Hamiltonian~\eqref{eq:H}, where the number of laser-induced excitons is given by $|b_L|^2 = (I/ \omega_L)\, \Gamma_0 /[(\omega_L - \omega_x)^2+\Gamma_x^2]$, $I$ is the laser power, $\Delta=\omega_L-\omega_x$ is the laser detunig, $\Gamma_0$ and $\Gamma_x$ are the radiative and total exciton decay rates, respectively. The presence of a exciton population enhances the exciton-phonon interaction leading to the modification of the exciton spectrum. This effect is described in the framework of Keldysh diagram technique~\cite{Keldysh1986} by the exciton self-energy correction
\begin{equation}\label{Sigma}
\Sigma(\omega) = |b_L|^2 \sum_k |g_k|^2 D_k(\omega-\omega_L),
\end{equation}
where $D_k^{(0)}(\Omega) = 2\Omega_k/[(\Omega+\rmi \Gamma_p)^2 -\Omega_k^2 ]$ is the retarded phonon Green's function, and $\Gamma_p$ is the phonon decay rate due to the unharmonicity and scattering.
We calculate  the real and imaginary parts of the self-energy, $\Sigma(\omega_L+\Omega) = \delta\Omega(\Omega) - \rmi \gamma(\Omega)$, separately and obtain
\begin{align}
&\gamma(\Omega) = \frac{|b_L|^2 |\Xi_k|^2 \Omega}{2\rho s^3 S} \,,\\
&\delta\Omega(\Omega)  = -\frac{|b_L|^2}{\rho s^2 S} \,{\rm V.p.}\int  \frac{ \Omega_{k'}^2|\Xi_{k'}|^2}{\Omega_{k'}^2-\Omega^2}  \frac{dk'}{2\pi}\,.
\end{align} 
The real part of self-energy $\delta\Omega$ describes a modification of the exciton resonance frequency. 
From now on we assume that it has been  already included in the exciton energy.
Contrary, the imaginary part $\gamma$ leads to the pumping-induced modification of exciton lifetime that is shown below to cause phonon amplification and attenuation.

\begin{figure}[t]
  \includegraphics[width=.85\columnwidth]{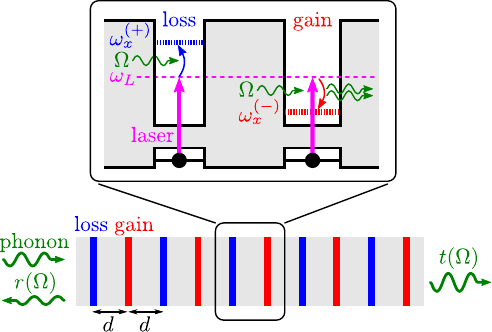}
 \caption{Laser-pumped multiple-quantum-well structure with two quantum wells in the unit cell that realizes the $\mathcal{PT}$ symmetry for phonons. }\label{fig:pic}
\end{figure}

The transmission $t(\Omega)$ and reflection $r(\Omega)$ coefficients of a phonon with the frequency $\Omega$ through the pumped QW read~\footnote{See Supplemental Material for the derivation of the reflection and transmission coefficients by means of the Keldysh diagram technique.},
\begin{align}
&t(\Omega) = \frac{\Delta^2 - (\Omega+\rmi\Gamma_x)^2}{\Delta^2 - (\Omega+\rmi\Gamma_x)^2+2\Delta\rmi\gamma(\Omega) }\,,\label{tk}\\
&r(\Omega) = 1-t(\Omega) \,,\label{rk}
\end{align}
where $\Delta = \omega_L - \omega_x$ is the detuning of the laser from the exciton resonance.
It follows from Eqs.~\eqref{tk}-\eqref{rk} that the phonon transmission and reflection coefficients have a pole at the frequency $\Omega \approx |\Delta| - \rmi (\Gamma_x - \gamma\, {\rm sign\,} \Delta)$, that shifts towards (away from) the real axis with increasing $\gamma$ for positive (negative) laser detuning. This effect is similar to the optomechanical heating (cooling)~\cite{Kippenberg2014} but utilizes the exciton resonance instead of cavity resonance.  The maximal reflectance is achieved at the resonance frequency $\Omega_0 = \sqrt{\Delta^2+\Gamma_x^2}$. Transmittance also has an extremum at $\Omega_0$, the resonant value of the transmission coefficient reads $t(\Omega) \approx 1 + (\gamma \Delta)/(\Gamma_x\Omega_0)$. Hence, the transmitted phonon is amplified if $\Delta>0$ and attenuated if $\Delta<0$.

We estimate the effect of phonon amplification using the state-of-the-art GaAs QW parameters~\cite{Jusserand2015,Trifonov2015}.
We take $\Xi_k \approx  10\,$eV, $\Gamma_x=0.1\,$meV and consider $120$\,GHz phonons. Then for an exciton density realizable in experiment $|b_L|^2/S = 10^{10}\,$cm$^{-2}$~\cite{kavbamalas} we obtain $\gamma \approx 1 \mu$eV. While this corresponds only to a 1\% amplification by a single QW, in  MQW 
structures with hundreds of QWs~\cite{Prineas2000}, amplification by several times is expected. 
Methods of acoustic transmission spectroscopy applicable in this context have been already developed~\cite{Huynh2006}.

\begin{figure}
  \includegraphics[width=.99\columnwidth]{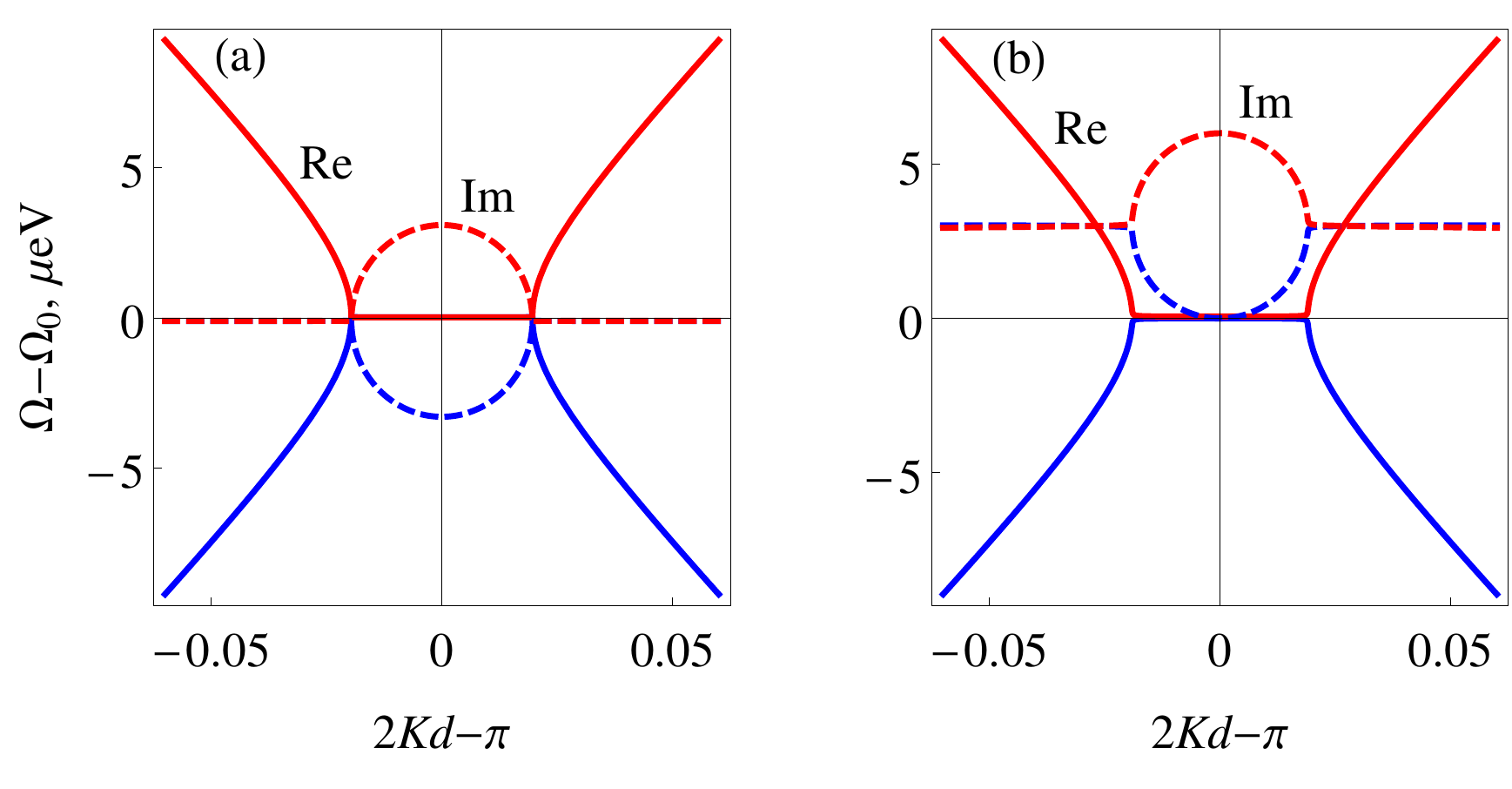}
\caption{Phonon dispersion for (a) the $\mathcal{PT}$-symmetric structure and (b) the structure with amplifying QWs only: dependence of the real (solid) and imaginary (dashed) parts of the phonon frequency on the phonon Bloch wave vector. The parameters are $\delta\omega_x = 0.5$\,meV, $\Gamma_x = 0.1$\,meV, $\gamma(\Omega_0)=1\,\mu$eV, $\Gamma_p = 0$. The inter-well distance $d=10$\,nm is tuned to satisfy the Bragg condition at the frequency $\Omega_0$. The acoustic contrast is neglected.
}\label{fig:disp}
\end{figure}
\begin{figure*}[t!]
  \includegraphics[width=.99\textwidth]{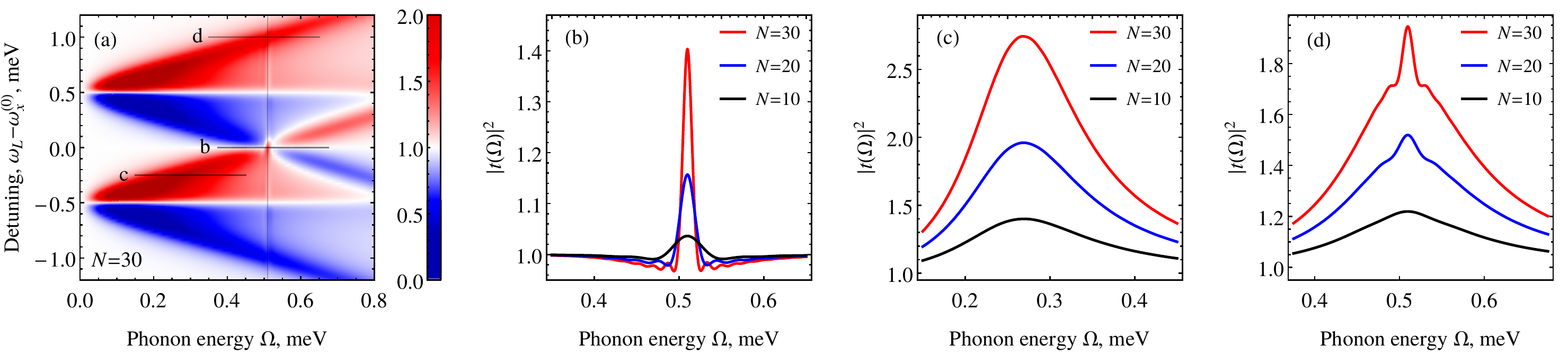}
 \caption{(a) Map of transmittance as a function of the phonon frequency $\Omega$ and the laser energy detuning $\omega_L-\omega_x^{(0)}$ for a structure with $N=30$ periods. Vertical line indicates the  Bragg resonance frequency. (b)-(d) Cross-sections of the map at fixed laser detunings, indicated by horizontal dashed lines in panel (a).   Panel (b) corresponds to the laser energy that realizes the $\mathcal{PT}$ symmetry.  The parameters are the same as those for Fig.~\ref{fig:disp}. }\label{fig:p4}
\end{figure*}

{\it Resonant $\mathcal{PT}$-symmetric phononic crystal.}---We have shown above that the sound amplification/attenuation is controlled by the laser detuning  from the exciton resonance. Now we consider a periodic layered structure consisting of QWs separated by barriers of width $d$, see Fig.~\ref{fig:pic}. The unit cell of the structure contains two QWs of slightly different parameters so that their exciton resonance energies are distinct, $\omega_x^{(\pm)}=\omega_x^{(0)} \pm \delta\omega_x$.  When the structure is pumped with a laser of the frequency $\omega_L$ that lies between the two exciton resonances, a half of QWs will amplify the sound while the other half will attenuate it. We focus on the case when the laser frequency $\omega_L$ is equal to $\omega_x^{(0)}$ and the sound gain and loss in QWs of the two types have the same magnitude, hence the acoustic $\mathcal{PT}$ symmetry is realized.

In order to calculate the phonon transmission coefficient through the structure we use the transfer matrix technique for acoustic waves~\cite{Ivchenko2005}.
For each structure layer, a QW or a barrier, we build a $2\times 2$ transfer matrix $\hat{T}$ linking the amplitudes of the right- and left-going acoustic waves (denoted by $+$ and $-$, respectively) at the right layer edge  with those at the left one,
\begin{equation}\label{eq:T1}
\hat{T}(\Omega) = \frac{1}{t(\Omega)}
\begin{bmatrix}
t^2(\Omega) - r^2(\Omega) \; & r(\Omega) \\ - r(\Omega) & 1
\end{bmatrix}
\:.
\end{equation}
Here, the single layer transmission and reflection coefficients $t$ and $r$ are given by Eqs.~\eqref{tk}-\eqref{rk} for a QW, while for an interwell barrier they are equal to $r=0$ and $t=\e^{\rmi k d} $, $k= (\Omega + \rmi \Gamma_p)/s$. The transfer matrix $\hat T_1$ for a structure period is a product of the transfer matrices for two QWs with $\Delta=\pm \delta\omega_x$ and two barriers of width $d$. 
By calculating the eigenvalues $\e^{2\rmi Kd}$ of the matrix $\hat{T}_1$ we obtain the dispersion equation
\begin{align}\label{disp}
\sin^2 Kd = \left[ 1 - \chi^2(\Omega) \right]\sin^2 \frac{\Omega+\rmi\Gamma_p}sd \,,
\end{align}
where $\chi(\Omega) = 2\rmi\gamma \delta\omega_x/[(\Omega+\rmi\Gamma_x)^2-\delta\omega_x^2]$. Equation~\eqref{disp} determines the relation between the Bloch wave vector of an acoustic wave $K$ and its frequency $\Omega$. 
First, we neglect  the finite phonon lifetime $\Gamma_p$.
If we choose $\Omega$ to be real, we will get from Eq.~\eqref{disp} the value of $K(\Omega)$ that has, in general, a non-zero imaginary part. The only exception is the frequency $\Omega_0 =  \sqrt{\delta\omega_x^2+\Gamma_x^2}$, where $\chi(\Omega_0)$ turns real, so Bloch wave vector $K(\Omega_0)$ is also real and acoustic wave does not decay nor grow in space. 
The frequency $\Omega_0$ is the one where  the exact $\mathcal{PT}$ symmetry for sound is realized.
Note that due to the causality principle, the exact $\mathcal{PT}$ symmetry can be realized at isolated frequency points only~\cite{Zyablovsky_2014}, in our case at the frequency $\Omega_0$. The discussed below effects of $\mathcal{PT}$ symmetry are still valid in a certain vicinity of this frequency, where the deviation from  $\mathcal{PT}$ symmetry is negligible.

{\it $\mathcal{PT}$ symmetry breaking for Bragg phonons.}---Of particular interest is the case when the frequency of $\mathcal{PT}$ symmetry matches the  Bragg resonance condition, $\Omega_0/s = \pi/2d$. In what follows we focus on this case.  
Using the Taylor expansion for Eq.~\eqref{disp} in the vicinity of $\Omega=\Omega_0$ and $K=\pi/2d$ we obtain the dispersion law of acoustic Bloch modes
\begin{align}\label{dispB}
\Omega(K) = \Omega_0  \pm \sqrt{ s^2 (K - \pi/2d)^2 - G^2}\,,
\end{align}
where $G = (s/d)\chi(\Omega_0) =  2\gamma\delta\omega_x/\pi\Gamma_x$ is the maximal mode decay/gain rate. The dispersion law of the real and imaginary parts of the frequency $\Omega(K)$ of the acoustic Bloch mode  in the $\mathcal{PT}$-symmetric structure is shown in Fig.~\ref{fig:disp}a. It follows from Eq.~\eqref{dispB} and Fig.~\ref{fig:disp} that the frequencies $\Omega(K)$ corresponding to the Bloch states with wave vector far from the Brillouin zone edge, $|K-\pi/2d|>G/s$, are real, which is a consequence of $\mathcal{PT}$ symmetry. On the other hand, the frequencies of the two Bragg acoustic modes with the wave vector $|K-\pi/2d|<G/s$ have non-zero imaginary parts and are complex conjugate to each other. This can be interpreted as $\mathcal{PT}$ symmetry breaking for these modes: Indeed, the standing acoustic wave with the wave vector $K=\pi/2d$ matching the Bragg condition can either have the deformation nodes in the amplifying QWs and  the antinodes in the attenuating QWs, or vice versa, resulting in the net gain or loss, respectively. 

For comparison we consider the structure where all attenuating QWs are removed, i.e. an array of amplifying QWs with the inter-well distance $2d$. The sound dispersion in such structure is shown in Fig.~\ref{fig:disp}b. In that case the frequencies of all Bloch modes have positive imaginary part. As a result,  many acoustic modes are amplified simultaneously, in contrast to the $\mathcal{PT}$-symmetric structure, where the modes only within a narrow range of wave vectors are amplified.

{\it Highly selective sound amplification.}---The sound transmission coefficient through the structure with $N$ periods is readily expressed via the corresponding transfer matrix element as $t_N(\Omega) = 1/[\hat T_1^N(\Omega)]_{--}$. 
Figure~\ref{fig:p4}a shows the map of the $\mathcal{PT}$-symmetric structure transmittance vs the phonon frequency and the laser detuning. The areas of amplification (red color) of transmitted sound lie above of either of two resonances, while the areas of attenuation (blue) lie below them. In agreement with Eq.~\ref{tk}, maximal amplification/attenuation occurs when $\Omega = |\omega_L - \omega_x^{(\pm)}|$. The areas of amplification and attenuation intersect at $\omega_L = \omega_x^{(0)}$, which is a signature of acoustic $\mathcal{PT}$ symmetry. The cross-section of the map showing the dependence of transmittance on phonon frequency in this case is shown in Fig.~\ref{fig:p4}b. One can see that the transmittance is close to unity at all phonon frequencies except for a narrow frequency region around the Bragg resonance. The former is the consequence of the $\mathcal{PT}$ symmetry while the latter is due to  the $\mathcal{PT}$ symmetry breaking for the Bragg phonons discussed above.  
Transmission spectrum in the vicinity of the Bragg resonance is  described by
\begin{align}\label{tNB}
|t_N(\Omega_0+\omega)|^2 = 1 + \frac{G^2 \sin^2 [\sqrt{\omega^2+G^2}L/s]}{\omega^2 + G^2 \cos^2 [ \sqrt{\omega^2+G^2}L/s]} \,,
\end{align}
where $L=2dN$ is the structure length.
It follows from Eq.~\eqref{tNB} and Fig.~\ref{fig:p4}b that the width of the frequency region where the sound is amplified shrinks as $ 2\sqrt{(\Omega_0/N)^2-G^2}$ with the increase of the structure length.

The highly selective sound amplification is a hallmark of $\mathcal{PT}$-symmetric system. For comparison we consider the case when the laser is detuned from $\omega_x^{(0)}$ and no $\mathcal{PT}$ symmetry is present. Shown in Fig.~\ref{fig:p4}c is the sound transmittance for the case when laser frequency lies slightly above the exciton resonance $\omega_x^{(-)}$. In that case, the sound is amplified in a much broader frequency range of the width $\Gamma_x$ that does not change with the increase of the structure length.  Figure~\ref{fig:p4}d corresponds to the laser frequency tuned in such way that the Bragg condition is satisfied only at the high-energy  exciton resonance, $\omega_L=\omega_x^{(+)}+\pi s/(2d)$. In that case the QWs with low-energy exciton are off-resonant, so the structure is equivalent to that with  all low-energy QWs  removed. Phonon dispersion in such structure suggests that the Bloch modes with positive imaginary part of the eigenfrequency exist in a wide range of wave vectors, even though there is some boost at the Bragg resonance, see Fig.~\ref{fig:disp}b. Concomitantly, in the transmission spectrum Fig.~\ref{fig:p4}d 
both the narrow peak and the wide background are present, so the resulting amplification selectivity is low.

\begin{figure}
  \includegraphics[width=.99\columnwidth]{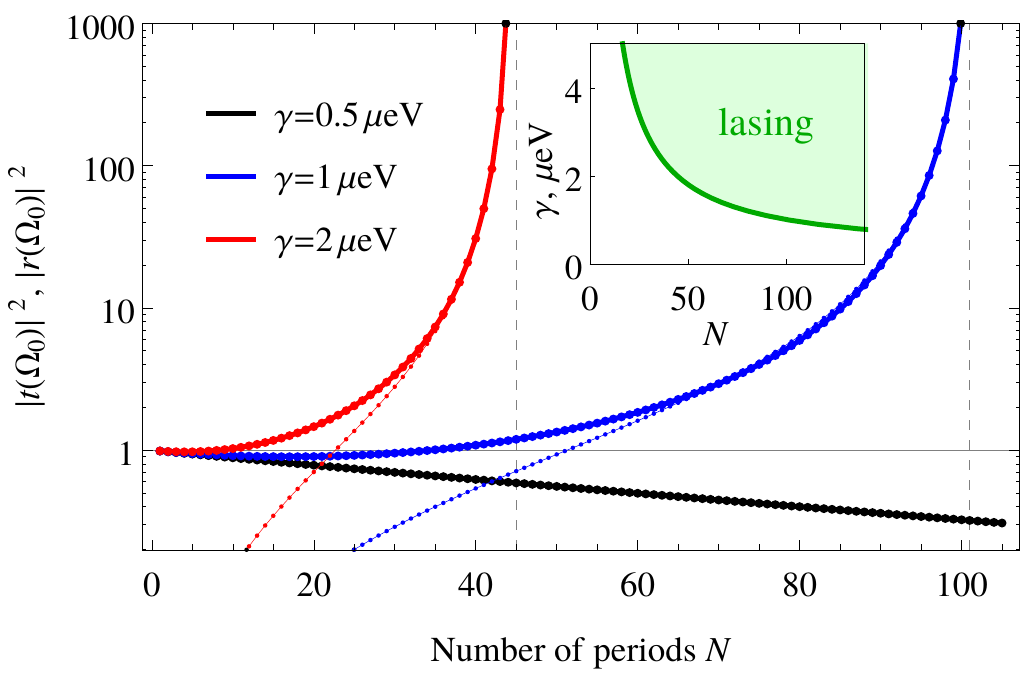}
 \caption{Dependence of the resonant  sound transmission (thick) and reflection (thin) coefficients on the number of periods in the structure $N$ plotted for different powers of pump laser, characterized by the value of $\gamma$. Vertical dashed lines indicate the acoustic lasing transition. Inset: Dependence of the critical value of $\gamma$ on the number of periods $N$. Calculation is made for $\Gamma_p = 1\,\mu$eV and other parameters are the same as for Fig.~\ref{fig:disp}.  
 }\label{fig:tMg}
\end{figure}

{\it Phonon lasing transition.}---Finally, we discuss the possible application of the proposed $\mathcal{PT}$-symmetric acoustic crystal to achieve distributed feedback single-mode phonon lasing. With increase of the structure length the amplification peak grows while its width decreases, see Fig.~\ref{fig:p4}b. 
We plot in Fig.~\ref{fig:tMg} by thick  solid curves the dependence of the transmission maximum $|t(\Omega_0)|^2$ on the number of periods in the structure.
The curves are calculated for  the different values of $\gamma$ that are proportional to the pump laser intensities, and for finite phonon decay rate $\Gamma_p=1\,\mu$eV~\cite{Rozas2009}.
At low pump rate the transmission coefficient tends to zero when $N \to \infty$, see the black curve in Fig.~\ref{fig:tMg}. 
The behavior changes drastically when the value of Bragg mode amplification $G$ surpasses the phonon decay rate $\Gamma_p$. This takes place for $\gamma > \pi\Gamma_x\Gamma_p/2\delta\omega_x$ and means that the infinite structure becomes unstable.
Consequently, the dependence of the transmission coefficient on the structure length diverges at a certain number of periods, see blue and red curves in Fig.~\ref{fig:tMg}. The diverging transmission coefficient indicates that a complex eigenfrequency of the structure has reached the real axis and further increase of the structure length or the laser power will lead to this pole acquiring a positive imaginary part, meaning the phonon lasing transition. The critical number of periods is given by
\begin{equation}
 N_c = \frac{1}{\pi} \frac{\Omega_0}{\sqrt{\smash[b]{G^2-\Gamma_p^2}}} \left( \frac{\pi}{2} + {\rm arcsin\,} \frac{\Gamma_p}{G}\right) \,,
\end{equation}
and decreases with the increase of the laser power. The dependence of the critical value of $\gamma$ on the number of periods in the structure is shown in the inset of Fig.~\ref{fig:tMg}. The value of $\gamma = 1\,\mu$eV corresponds approximately to the pump laser intensity 1\,mW per $\mu$m$^2$ cross-section.

Even if the structure itself is amplifying but not yet lasing, one can attain the lasing by embedding it into an acoustic cavity~\cite{Trigo2002}. The simplest approach is to place the considered above resonant acoustic crystal near an interface with vacuum. Assuming perfect phonon reflection from the interface, and tuning the distance from the interface to the first amplifying QW to $\lambda/4$, we will get lasing as soon as $|r(\Omega_0)|>1$.  Shown in Fig.~\ref{fig:tMg} by thin curves is the dependence of the reflection coefficient on the structure length. One can see that $|r(\Omega_0)|$ reaches unity in the structures that are twice shorter than critical. This is because the mirror reflection from the interface makes the structure efficiently twice longer, consequently reducing the critical length of lasing by the factor of 2.

To summarize, we propose to use laser-pumped quantum wells for phonon manipulation at the nanoscale level. Phonon amplification and attenuation by a quantum well with realistic parameters can be controlled by the laser detuning and intensity. We show how to combine quantum wells with red- and blue-detuned exciton resonances in the same structure in order to realize the acoustic $\mathcal{PT}$ symmetry. Such resonant acoustic crystal turns out to be a highly-selective sound amplifier, which can be used to achieve the $\mathcal{PT}$-symmetric single-mode phonon lasing.

\paragraph*{Acknowledgments.} 
This work was supported by
the RFBR and the Foundation ``Dynasty.'' A.N.P. and A.V.P.
also acknowledge support by the Russian President Grants No.
MK-8500.2016.2 and SP-2912.2016.5, respectively.

\bibliography{pt.bib}

\begin{thebibliography}{37}%
\makeatletter
\providecommand \@ifxundefined [1]{%
 \@ifx{#1\undefined}
}%
\providecommand \@ifnum [1]{%
 \ifnum #1\expandafter \@firstoftwo
 \else \expandafter \@secondoftwo
 \fi
}%
\providecommand \@ifx [1]{%
 \ifx #1\expandafter \@firstoftwo
 \else \expandafter \@secondoftwo
 \fi
}%
\providecommand \natexlab [1]{#1}%
\providecommand \enquote  [1]{``#1''}%
\providecommand \bibnamefont  [1]{#1}%
\providecommand \bibfnamefont [1]{#1}%
\providecommand \citenamefont [1]{#1}%
\providecommand \href@noop [0]{\@secondoftwo}%
\providecommand \href [0]{\begingroup \@sanitize@url \@href}%
\providecommand \@href[1]{\@@startlink{#1}\@@href}%
\providecommand \@@href[1]{\endgroup#1\@@endlink}%
\providecommand \@sanitize@url [0]{\catcode `\\12\catcode `\$12\catcode
  `\&12\catcode `\#12\catcode `\^12\catcode `\_12\catcode `\%12\relax}%
\providecommand \@@startlink[1]{}%
\providecommand \@@endlink[0]{}%
\providecommand \url  [0]{\begingroup\@sanitize@url \@url }%
\providecommand \@url [1]{\endgroup\@href {#1}{\urlprefix }}%
\providecommand \urlprefix  [0]{URL }%
\providecommand \Eprint [0]{\href }%
\providecommand \doibase [0]{http://dx.doi.org/}%
\providecommand \selectlanguage [0]{\@gobble}%
\providecommand \bibinfo  [0]{\@secondoftwo}%
\providecommand \bibfield  [0]{\@secondoftwo}%
\providecommand \translation [1]{[#1]}%
\providecommand \BibitemOpen [0]{}%
\providecommand \bibitemStop [0]{}%
\providecommand \bibitemNoStop [0]{.\EOS\space}%
\providecommand \EOS [0]{\spacefactor3000\relax}%
\providecommand \BibitemShut  [1]{\csname bibitem#1\endcsname}%
\let\auto@bib@innerbib\@empty
\bibitem [{\citenamefont {Bender}\ and\ \citenamefont
  {Boettcher}(1998)}]{Bender1998}%
  \BibitemOpen
  \bibfield  {author} {\bibinfo {author} {\bibfnamefont {C.~M.}\ \bibnamefont
  {Bender}}\ and\ \bibinfo {author} {\bibfnamefont {S.}~\bibnamefont
  {Boettcher}},\ }\bibfield  {title} {\enquote {\bibinfo {title} {Real spectra
  in non-hermitian {H}amiltonians having $\mathcal{PT}$ symmetry},}\ }\href
  {\doibase 10.1103/PhysRevLett.80.5243} {\bibfield  {journal} {\bibinfo
  {journal} {Phys. Rev. Lett.}\ }\textbf {\bibinfo {volume} {80}},\ \bibinfo
  {pages} {5243--5246} (\bibinfo {year} {1998})}\BibitemShut {NoStop}%
\bibitem [{\citenamefont {Zyablovsky}\ \emph
  {et~al.}(2014{\natexlab{a}})\citenamefont {Zyablovsky}, \citenamefont
  {Vinogradov}, \citenamefont {Pukhov}, \citenamefont {Dorofeenko},\ and\
  \citenamefont {Lisyansky}}]{Zyablovsky2014ufn}%
  \BibitemOpen
  \bibfield  {author} {\bibinfo {author} {\bibfnamefont {A.~A.}\ \bibnamefont
  {Zyablovsky}}, \bibinfo {author} {\bibfnamefont {A.~P.}\ \bibnamefont
  {Vinogradov}}, \bibinfo {author} {\bibfnamefont {A.~A.}\ \bibnamefont
  {Pukhov}}, \bibinfo {author} {\bibfnamefont {A.~V.}\ \bibnamefont
  {Dorofeenko}}, \ and\ \bibinfo {author} {\bibfnamefont {A.~A.}\ \bibnamefont
  {Lisyansky}},\ }\bibfield  {title} {\enquote {\bibinfo {title} {{PT}-symmetry
  in optics},}\ }\href {\doibase 10.3367/ufnr.0184.201411b.1177} {\bibfield
  {journal} {\bibinfo  {journal} {Usp. Fiz. Nauk}\ }\textbf {\bibinfo {volume}
  {184}},\ \bibinfo {pages} {1177--1198} (\bibinfo {year}
  {2014}{\natexlab{a}})}\BibitemShut {NoStop}%
\bibitem [{\citenamefont {{Suchkov}}\ \emph {et~al.}(2015)\citenamefont
  {{Suchkov}}, \citenamefont {{Sukhorukov}}, \citenamefont {{Huang}},
  \citenamefont {{Dmitriev}}, \citenamefont {{Lee}},\ and\ \citenamefont
  {{Kivshar}}}]{Suchkov2015}%
  \BibitemOpen
  \bibfield  {author} {\bibinfo {author} {\bibfnamefont {S.~V.}\ \bibnamefont
  {{Suchkov}}}, \bibinfo {author} {\bibfnamefont {A.~A.}\ \bibnamefont
  {{Sukhorukov}}}, \bibinfo {author} {\bibfnamefont {J.}~\bibnamefont
  {{Huang}}}, \bibinfo {author} {\bibfnamefont {S.~V.}\ \bibnamefont
  {{Dmitriev}}}, \bibinfo {author} {\bibfnamefont {C.}~\bibnamefont {{Lee}}}, \
  and\ \bibinfo {author} {\bibfnamefont {Y.~S.}\ \bibnamefont {{Kivshar}}},\
  }\bibfield  {title} {\enquote {\bibinfo {title} {{Nonlinear switching and
  solitons in PT-symmetric photonic systems}},}\ }\href@noop {} {\bibfield
  {journal} {\bibinfo  {journal} {ArXiv e-prints}\ } (\bibinfo {year}
  {2015})},\ \Eprint {http://arxiv.org/abs/1509.03378} {arXiv:1509.03378
  [physics.optics]} \BibitemShut {NoStop}%
\bibitem [{\citenamefont {{Konotop}}\ \emph {et~al.}(2016)\citenamefont
  {{Konotop}}, \citenamefont {{Yang}},\ and\ \citenamefont
  {{Zezyulin}}}]{Konotop2016}%
  \BibitemOpen
  \bibfield  {author} {\bibinfo {author} {\bibfnamefont {V.~V.}\ \bibnamefont
  {{Konotop}}}, \bibinfo {author} {\bibfnamefont {J.}~\bibnamefont {{Yang}}}, \
  and\ \bibinfo {author} {\bibfnamefont {D.~A.}\ \bibnamefont {{Zezyulin}}},\
  }\bibfield  {title} {\enquote {\bibinfo {title} {{Nonlinear waves in $\cal
  PT$-symmetric systems}},}\ }\href@noop {} {\bibfield  {journal} {\bibinfo
  {journal} {ArXiv e-prints}\ } (\bibinfo {year} {2016})},\ \Eprint
  {http://arxiv.org/abs/1603.06826} {arXiv:1603.06826 [nlin.PS]} \BibitemShut
  {NoStop}%
\bibitem [{\citenamefont {R\"{u}ter}\ \emph {et~al.}(2010)\citenamefont
  {R\"{u}ter}, \citenamefont {Makris}, \citenamefont {El-Ganainy},
  \citenamefont {Christodoulides}, \citenamefont {Segev},\ and\ \citenamefont
  {Kip}}]{Ruter2010}%
  \BibitemOpen
  \bibfield  {author} {\bibinfo {author} {\bibfnamefont {C.~E.}\ \bibnamefont
  {R\"{u}ter}}, \bibinfo {author} {\bibfnamefont {K.~G.}\ \bibnamefont
  {Makris}}, \bibinfo {author} {\bibfnamefont {R.}~\bibnamefont {El-Ganainy}},
  \bibinfo {author} {\bibfnamefont {D.~N.}\ \bibnamefont {Christodoulides}},
  \bibinfo {author} {\bibfnamefont {M.}~\bibnamefont {Segev}}, \ and\ \bibinfo
  {author} {\bibfnamefont {D.}~\bibnamefont {Kip}},\ }\bibfield  {title}
  {\enquote {\bibinfo {title} {Observation of parity{\textendash}time symmetry
  in optics},}\ }\href {\doibase 10.1038/nphys1515} {\bibfield  {journal}
  {\bibinfo  {journal} {Nat. Phys.}\ }\textbf {\bibinfo {volume} {6}},\
  \bibinfo {pages} {192--195} (\bibinfo {year} {2010})}\BibitemShut {NoStop}%
\bibitem [{\citenamefont {Hodaei}\ \emph {et~al.}(2014)\citenamefont {Hodaei},
  \citenamefont {Miri}, \citenamefont {Heinrich}, \citenamefont
  {Christodoulides},\ and\ \citenamefont {Khajavikhan}}]{Hodaei2014}%
  \BibitemOpen
  \bibfield  {author} {\bibinfo {author} {\bibfnamefont {H.}~\bibnamefont
  {Hodaei}}, \bibinfo {author} {\bibfnamefont {M.-A.}\ \bibnamefont {Miri}},
  \bibinfo {author} {\bibfnamefont {M.}~\bibnamefont {Heinrich}}, \bibinfo
  {author} {\bibfnamefont {D.~N.}\ \bibnamefont {Christodoulides}}, \ and\
  \bibinfo {author} {\bibfnamefont {M.}~\bibnamefont {Khajavikhan}},\
  }\bibfield  {title} {\enquote {\bibinfo {title} {Parity-time-symmetric
  microring lasers},}\ }\href {\doibase 10.1126/science.1258480} {\bibfield
  {journal} {\bibinfo  {journal} {Science}\ }\textbf {\bibinfo {volume}
  {346}},\ \bibinfo {pages} {975--978} (\bibinfo {year} {2014})}\BibitemShut
  {NoStop}%
\bibitem [{\citenamefont {Peng}\ \emph {et~al.}(2014)\citenamefont {Peng},
  \citenamefont {\"{O}zdemir}, \citenamefont {Lei}, \citenamefont {Monifi},
  \citenamefont {Gianfreda}, \citenamefont {Long}, \citenamefont {Fan},
  \citenamefont {Nori}, \citenamefont {Bender},\ and\ \citenamefont
  {Yang}}]{Peng2014b}%
  \BibitemOpen
  \bibfield  {author} {\bibinfo {author} {\bibfnamefont {B.}~\bibnamefont
  {Peng}}, \bibinfo {author} {\bibfnamefont {{\c{S}}.~K.}\ \bibnamefont
  {\"{O}zdemir}}, \bibinfo {author} {\bibfnamefont {F.}~\bibnamefont {Lei}},
  \bibinfo {author} {\bibfnamefont {F.}~\bibnamefont {Monifi}}, \bibinfo
  {author} {\bibfnamefont {M.}~\bibnamefont {Gianfreda}}, \bibinfo {author}
  {\bibfnamefont {G.~L.}\ \bibnamefont {Long}}, \bibinfo {author}
  {\bibfnamefont {S.}~\bibnamefont {Fan}}, \bibinfo {author} {\bibfnamefont
  {F.}~\bibnamefont {Nori}}, \bibinfo {author} {\bibfnamefont {C.~M.}\
  \bibnamefont {Bender}}, \ and\ \bibinfo {author} {\bibfnamefont
  {L.}~\bibnamefont {Yang}},\ }\bibfield  {title} {\enquote {\bibinfo {title}
  {Parity{\textendash}time-symmetric whispering-gallery microcavities},}\
  }\href {\doibase 10.1038/nphys2927} {\bibfield  {journal} {\bibinfo
  {journal} {Nat. Phys.}\ }\textbf {\bibinfo {volume} {10}},\ \bibinfo {pages}
  {394--398} (\bibinfo {year} {2014})}\BibitemShut {NoStop}%
\bibitem [{\citenamefont {Lin}\ \emph {et~al.}(2011)\citenamefont {Lin},
  \citenamefont {Ramezani}, \citenamefont {Eichelkraut}, \citenamefont
  {Kottos}, \citenamefont {Cao},\ and\ \citenamefont
  {Christodoulides}}]{Lin2011}%
  \BibitemOpen
  \bibfield  {author} {\bibinfo {author} {\bibfnamefont {Z.}~\bibnamefont
  {Lin}}, \bibinfo {author} {\bibfnamefont {H.}~\bibnamefont {Ramezani}},
  \bibinfo {author} {\bibfnamefont {T.}~\bibnamefont {Eichelkraut}}, \bibinfo
  {author} {\bibfnamefont {T.}~\bibnamefont {Kottos}}, \bibinfo {author}
  {\bibfnamefont {H.}~\bibnamefont {Cao}}, \ and\ \bibinfo {author}
  {\bibfnamefont {D.~N.}\ \bibnamefont {Christodoulides}},\ }\bibfield  {title}
  {\enquote {\bibinfo {title} {Unidirectional invisibility induced by
  $\mathcal{P}\mathcal{T}$-symmetric periodic structures},}\ }\href {\doibase
  10.1103/PhysRevLett.106.213901} {\bibfield  {journal} {\bibinfo  {journal}
  {Phys. Rev. Lett.}\ }\textbf {\bibinfo {volume} {106}},\ \bibinfo {pages}
  {213901} (\bibinfo {year} {2011})}\BibitemShut {NoStop}%
\bibitem [{\citenamefont {Feng}\ \emph {et~al.}(2012)\citenamefont {Feng},
  \citenamefont {Xu}, \citenamefont {Fegadolli}, \citenamefont {Lu},
  \citenamefont {Oliveira}, \citenamefont {Almeida}, \citenamefont {Chen},\
  and\ \citenamefont {Scherer}}]{Feng2012}%
  \BibitemOpen
  \bibfield  {author} {\bibinfo {author} {\bibfnamefont {L.}~\bibnamefont
  {Feng}}, \bibinfo {author} {\bibfnamefont {Y.-L.}\ \bibnamefont {Xu}},
  \bibinfo {author} {\bibfnamefont {W.~S.}\ \bibnamefont {Fegadolli}}, \bibinfo
  {author} {\bibfnamefont {M.-H.}\ \bibnamefont {Lu}}, \bibinfo {author}
  {\bibfnamefont {J.~E.~B.}\ \bibnamefont {Oliveira}}, \bibinfo {author}
  {\bibfnamefont {V.~R.}\ \bibnamefont {Almeida}}, \bibinfo {author}
  {\bibfnamefont {Y.-F.}\ \bibnamefont {Chen}}, \ and\ \bibinfo {author}
  {\bibfnamefont {A.}~\bibnamefont {Scherer}},\ }\bibfield  {title} {\enquote
  {\bibinfo {title} {Experimental demonstration of a unidirectional
  reflectionless parity-time metamaterial at optical~frequencies},}\ }\href
  {\doibase 10.1038/nmat3495} {\bibfield  {journal} {\bibinfo  {journal} {Nat.
  Mater.}\ }\textbf {\bibinfo {volume} {12}},\ \bibinfo {pages} {108--113}
  (\bibinfo {year} {2012})}\BibitemShut {NoStop}%
\bibitem [{\citenamefont {Regensburger}\ \emph {et~al.}(2012)\citenamefont
  {Regensburger}, \citenamefont {Bersch}, \citenamefont {Miri}, \citenamefont
  {Onishchukov}, \citenamefont {Christodoulides},\ and\ \citenamefont
  {Peschel}}]{Regensburger2012}%
  \BibitemOpen
  \bibfield  {author} {\bibinfo {author} {\bibfnamefont {A.}~\bibnamefont
  {Regensburger}}, \bibinfo {author} {\bibfnamefont {C.}~\bibnamefont
  {Bersch}}, \bibinfo {author} {\bibfnamefont {M.-A.}\ \bibnamefont {Miri}},
  \bibinfo {author} {\bibfnamefont {G.}~\bibnamefont {Onishchukov}}, \bibinfo
  {author} {\bibfnamefont {D.~N.}\ \bibnamefont {Christodoulides}}, \ and\
  \bibinfo {author} {\bibfnamefont {U.}~\bibnamefont {Peschel}},\ }\bibfield
  {title} {\enquote {\bibinfo {title} {Parity{\textendash}time synthetic
  photonic lattices},}\ }\href {\doibase 10.1038/nature11298} {\bibfield
  {journal} {\bibinfo  {journal} {Nature}\ }\textbf {\bibinfo {volume} {488}},\
  \bibinfo {pages} {167--171} (\bibinfo {year} {2012})}\BibitemShut {NoStop}%
\bibitem [{\citenamefont {Chong}\ \emph {et~al.}(2011)\citenamefont {Chong},
  \citenamefont {Ge},\ and\ \citenamefont {Stone}}]{Stone2011}%
  \BibitemOpen
  \bibfield  {author} {\bibinfo {author} {\bibfnamefont {Y.~D.}\ \bibnamefont
  {Chong}}, \bibinfo {author} {\bibfnamefont {Li}~\bibnamefont {Ge}}, \ and\
  \bibinfo {author} {\bibfnamefont {A.~Douglas}\ \bibnamefont {Stone}},\
  }\bibfield  {title} {\enquote {\bibinfo {title}
  {$\mathcal{P}\mathcal{T}$-symmetry breaking and laser-absorber modes in
  optical scattering systems},}\ }\href {\doibase
  10.1103/PhysRevLett.106.093902} {\bibfield  {journal} {\bibinfo  {journal}
  {Phys. Rev. Lett.}\ }\textbf {\bibinfo {volume} {106}},\ \bibinfo {pages}
  {093902} (\bibinfo {year} {2011})}\BibitemShut {NoStop}%
\bibitem [{\citenamefont {Feng}\ \emph {et~al.}(2014)\citenamefont {Feng},
  \citenamefont {Wong}, \citenamefont {Ma}, \citenamefont {Wang},\ and\
  \citenamefont {Zhang}}]{Feng2014}%
  \BibitemOpen
  \bibfield  {author} {\bibinfo {author} {\bibfnamefont {L.}~\bibnamefont
  {Feng}}, \bibinfo {author} {\bibfnamefont {Z.~J.}\ \bibnamefont {Wong}},
  \bibinfo {author} {\bibfnamefont {R.-M.}\ \bibnamefont {Ma}}, \bibinfo
  {author} {\bibfnamefont {Y.}~\bibnamefont {Wang}}, \ and\ \bibinfo {author}
  {\bibfnamefont {X.}~\bibnamefont {Zhang}},\ }\bibfield  {title} {\enquote
  {\bibinfo {title} {Single-mode laser by parity-time symmetry breaking},}\
  }\href {\doibase 10.1126/science.1258479} {\bibfield  {journal} {\bibinfo
  {journal} {Science}\ }\textbf {\bibinfo {volume} {346}},\ \bibinfo {pages}
  {972--975} (\bibinfo {year} {2014})}\BibitemShut {NoStop}%
\bibitem [{\citenamefont {Zhu}\ \emph {et~al.}(2014)\citenamefont {Zhu},
  \citenamefont {Ramezani}, \citenamefont {Shi}, \citenamefont {Zhu},\ and\
  \citenamefont {Zhang}}]{Zhu2014}%
  \BibitemOpen
  \bibfield  {author} {\bibinfo {author} {\bibfnamefont {X.}~\bibnamefont
  {Zhu}}, \bibinfo {author} {\bibfnamefont {H.}~\bibnamefont {Ramezani}},
  \bibinfo {author} {\bibfnamefont {C.}~\bibnamefont {Shi}}, \bibinfo {author}
  {\bibfnamefont {J.}~\bibnamefont {Zhu}}, \ and\ \bibinfo {author}
  {\bibfnamefont {X.}~\bibnamefont {Zhang}},\ }\bibfield  {title} {\enquote
  {\bibinfo {title} {$\mathcal{PT}$-symmetric acoustics},}\ }\href {\doibase
  10.1103/physrevx.4.031042} {\bibfield  {journal} {\bibinfo  {journal} {Phys.
  Rev. X}\ }\textbf {\bibinfo {volume} {4}},\ \bibinfo {pages} {031042}
  (\bibinfo {year} {2014})}\BibitemShut {NoStop}%
\bibitem [{\citenamefont {Bender}\ \emph {et~al.}(2013)\citenamefont {Bender},
  \citenamefont {Berntson}, \citenamefont {Parker},\ and\ \citenamefont
  {Samuel}}]{Bender2013}%
  \BibitemOpen
  \bibfield  {author} {\bibinfo {author} {\bibfnamefont {C.~M.}\ \bibnamefont
  {Bender}}, \bibinfo {author} {\bibfnamefont {B.~K.}\ \bibnamefont
  {Berntson}}, \bibinfo {author} {\bibfnamefont {D.}~\bibnamefont {Parker}}, \
  and\ \bibinfo {author} {\bibfnamefont {E.}~\bibnamefont {Samuel}},\
  }\bibfield  {title} {\enquote {\bibinfo {title} {Observation of {PT} phase
  transition in a simple mechanical system},}\ }\href {\doibase
  10.1119/1.4789549} {\bibfield  {journal} {\bibinfo  {journal} {Am. J. Phys.}\
  }\textbf {\bibinfo {volume} {81}},\ \bibinfo {pages} {173} (\bibinfo {year}
  {2013})}\BibitemShut {NoStop}%
\bibitem [{\citenamefont {Christensen}\ \emph {et~al.}(2016)\citenamefont
  {Christensen}, \citenamefont {Willatzen}, \citenamefont {Velasco},\ and\
  \citenamefont {Lu}}]{Lu2016}%
  \BibitemOpen
  \bibfield  {author} {\bibinfo {author} {\bibfnamefont {J.}~\bibnamefont
  {Christensen}}, \bibinfo {author} {\bibfnamefont {M.}~\bibnamefont
  {Willatzen}}, \bibinfo {author} {\bibfnamefont {V.~R.}\ \bibnamefont
  {Velasco}}, \ and\ \bibinfo {author} {\bibfnamefont {M.-H.}\ \bibnamefont
  {Lu}},\ }\bibfield  {title} {\enquote {\bibinfo {title} {Parity-time
  synthetic phononic media},}\ }\href {\doibase 10.1103/PhysRevLett.116.207601}
  {\bibfield  {journal} {\bibinfo  {journal} {Phys. Rev. Lett.}\ }\textbf
  {\bibinfo {volume} {116}},\ \bibinfo {pages} {207601} (\bibinfo {year}
  {2016})}\BibitemShut {NoStop}%
\bibitem [{\citenamefont {Aspelmeyer}\ \emph {et~al.}(2014)\citenamefont
  {Aspelmeyer}, \citenamefont {Kippenberg},\ and\ \citenamefont
  {Marquardt}}]{Kippenberg2014}%
  \BibitemOpen
  \bibfield  {author} {\bibinfo {author} {\bibfnamefont {M.}~\bibnamefont
  {Aspelmeyer}}, \bibinfo {author} {\bibfnamefont {T.~J.}\ \bibnamefont
  {Kippenberg}}, \ and\ \bibinfo {author} {\bibfnamefont {F.}~\bibnamefont
  {Marquardt}},\ }\bibfield  {title} {\enquote {\bibinfo {title} {Cavity
  optomechanics},}\ }\href {\doibase 10.1103/RevModPhys.86.1391} {\bibfield
  {journal} {\bibinfo  {journal} {Rev. Mod. Phys.}\ }\textbf {\bibinfo {volume}
  {86}},\ \bibinfo {pages} {1391--1452} (\bibinfo {year} {2014})}\BibitemShut
  {NoStop}%
\bibitem [{\citenamefont {Xu}\ \emph {et~al.}(2015)\citenamefont {Xu},
  \citenamefont {Liu}, \citenamefont {Sun},\ and\ \citenamefont
  {Li}}]{Xu2015b}%
  \BibitemOpen
  \bibfield  {author} {\bibinfo {author} {\bibfnamefont {X.-W.}\ \bibnamefont
  {Xu}}, \bibinfo {author} {\bibfnamefont {Y.-x.}\ \bibnamefont {Liu}},
  \bibinfo {author} {\bibfnamefont {C.-P.}\ \bibnamefont {Sun}}, \ and\
  \bibinfo {author} {\bibfnamefont {Y.}~\bibnamefont {Li}},\ }\bibfield
  {title} {\enquote {\bibinfo {title} {Mechanical $\mathcal{PT}$ symmetry in
  coupled optomechanical systems},}\ }\href {\doibase
  10.1103/PhysRevA.92.013852} {\bibfield  {journal} {\bibinfo  {journal} {Phys.
  Rev. A}\ }\textbf {\bibinfo {volume} {92}},\ \bibinfo {pages} {013852}
  (\bibinfo {year} {2015})}\BibitemShut {NoStop}%
\bibitem [{\citenamefont {{Faist}}\ \emph {et~al.}(1994)\citenamefont
  {{Faist}}, \citenamefont {{Capasso}}, \citenamefont {{Sivco}}, \citenamefont
  {{Sirtori}}, \citenamefont {{Hutchinson}},\ and\ \citenamefont
  {{Cho}}}]{Capasso1994}%
  \BibitemOpen
  \bibfield  {author} {\bibinfo {author} {\bibfnamefont {J.}~\bibnamefont
  {{Faist}}}, \bibinfo {author} {\bibfnamefont {F.}~\bibnamefont {{Capasso}}},
  \bibinfo {author} {\bibfnamefont {D.~L.}\ \bibnamefont {{Sivco}}}, \bibinfo
  {author} {\bibfnamefont {C.}~\bibnamefont {{Sirtori}}}, \bibinfo {author}
  {\bibfnamefont {A.~L.}\ \bibnamefont {{Hutchinson}}}, \ and\ \bibinfo
  {author} {\bibfnamefont {A.~Y.}\ \bibnamefont {{Cho}}},\ }\bibfield  {title}
  {\enquote {\bibinfo {title} {{Quantum Cascade Laser}},}\ }\href {\doibase
  10.1126/science.264.5158.553} {\bibfield  {journal} {\bibinfo  {journal}
  {Science}\ }\textbf {\bibinfo {volume} {264}},\ \bibinfo {pages} {553--556}
  (\bibinfo {year} {1994})}\BibitemShut {NoStop}%
\bibitem [{\citenamefont {Kent}\ \emph {et~al.}(2006)\citenamefont {Kent},
  \citenamefont {Kini}, \citenamefont {Stanton}, \citenamefont {Henini},
  \citenamefont {Glavin}, \citenamefont {Kochelap},\ and\ \citenamefont
  {Linnik}}]{Kent2006}%
  \BibitemOpen
  \bibfield  {author} {\bibinfo {author} {\bibfnamefont {A.~J.}\ \bibnamefont
  {Kent}}, \bibinfo {author} {\bibfnamefont {R.~N.}\ \bibnamefont {Kini}},
  \bibinfo {author} {\bibfnamefont {N.~M.}\ \bibnamefont {Stanton}}, \bibinfo
  {author} {\bibfnamefont {M.}~\bibnamefont {Henini}}, \bibinfo {author}
  {\bibfnamefont {B.~A.}\ \bibnamefont {Glavin}}, \bibinfo {author}
  {\bibfnamefont {V.~A.}\ \bibnamefont {Kochelap}}, \ and\ \bibinfo {author}
  {\bibfnamefont {T.~L.}\ \bibnamefont {Linnik}},\ }\bibfield  {title}
  {\enquote {\bibinfo {title} {Acoustic phonon emission from a weakly coupled
  superlattice under vertical electron transport: Observation of phonon
  resonance},}\ }\href {\doibase 10.1103/PhysRevLett.96.215504} {\bibfield
  {journal} {\bibinfo  {journal} {Phys. Rev. Lett.}\ }\textbf {\bibinfo
  {volume} {96}},\ \bibinfo {pages} {215504} (\bibinfo {year}
  {2006})}\BibitemShut {NoStop}%
\bibitem [{\citenamefont {Maryam}\ \emph {et~al.}(2013)\citenamefont {Maryam},
  \citenamefont {Akimov}, \citenamefont {Campion},\ and\ \citenamefont
  {Kent}}]{Maryam2013}%
  \BibitemOpen
  \bibfield  {author} {\bibinfo {author} {\bibfnamefont {W.}~\bibnamefont
  {Maryam}}, \bibinfo {author} {\bibfnamefont {A.~V.}\ \bibnamefont {Akimov}},
  \bibinfo {author} {\bibfnamefont {R.~P.}\ \bibnamefont {Campion}}, \ and\
  \bibinfo {author} {\bibfnamefont {A.~J.}\ \bibnamefont {Kent}},\ }\bibfield
  {title} {\enquote {\bibinfo {title} {Dynamics of a vertical cavity quantum
  cascade phonon laser structure},}\ }\href {\doibase 10.1038/ncomms3184}
  {\bibfield  {journal} {\bibinfo  {journal} {Nat. Commun.}\ }\textbf {\bibinfo
  {volume} {4}} (\bibinfo {year} {2013}),\ 10.1038/ncomms3184}\BibitemShut
  {NoStop}%
\bibitem [{\citenamefont {Beardsley}\ \emph {et~al.}(2010)\citenamefont
  {Beardsley}, \citenamefont {Akimov}, \citenamefont {Henini},\ and\
  \citenamefont {Kent}}]{Beardsley2010}%
  \BibitemOpen
  \bibfield  {author} {\bibinfo {author} {\bibfnamefont {R.~P.}\ \bibnamefont
  {Beardsley}}, \bibinfo {author} {\bibfnamefont {A.~V.}\ \bibnamefont
  {Akimov}}, \bibinfo {author} {\bibfnamefont {M.}~\bibnamefont {Henini}}, \
  and\ \bibinfo {author} {\bibfnamefont {A.~J.}\ \bibnamefont {Kent}},\
  }\bibfield  {title} {\enquote {\bibinfo {title} {Coherent terahertz sound
  amplification and spectral line narrowing in a {S}tark ladder
  superlattice},}\ }\href {\doibase 10.1103/PhysRevLett.104.085501} {\bibfield
  {journal} {\bibinfo  {journal} {Phys. Rev. Lett.}\ }\textbf {\bibinfo
  {volume} {104}},\ \bibinfo {pages} {085501} (\bibinfo {year}
  {2010})}\BibitemShut {NoStop}%
\bibitem [{\citenamefont {Lanzillotti-Kimura}\ \emph
  {et~al.}(2007)\citenamefont {Lanzillotti-Kimura}, \citenamefont {Fainstein},
  \citenamefont {Huynh}, \citenamefont {Perrin}, \citenamefont {Jusserand},
  \citenamefont {Miard},\ and\ \citenamefont {Lema\^{\i}tre}}]{Kimura2007}%
  \BibitemOpen
  \bibfield  {author} {\bibinfo {author} {\bibfnamefont {N.~D.}\ \bibnamefont
  {Lanzillotti-Kimura}}, \bibinfo {author} {\bibfnamefont {A.}~\bibnamefont
  {Fainstein}}, \bibinfo {author} {\bibfnamefont {A.}~\bibnamefont {Huynh}},
  \bibinfo {author} {\bibfnamefont {B.}~\bibnamefont {Perrin}}, \bibinfo
  {author} {\bibfnamefont {B.}~\bibnamefont {Jusserand}}, \bibinfo {author}
  {\bibfnamefont {A.}~\bibnamefont {Miard}}, \ and\ \bibinfo {author}
  {\bibfnamefont {A.}~\bibnamefont {Lema\^{\i}tre}},\ }\bibfield  {title}
  {\enquote {\bibinfo {title} {Coherent generation of acoustic phonons in an
  optical microcavity},}\ }\href {\doibase 10.1103/PhysRevLett.99.217405}
  {\bibfield  {journal} {\bibinfo  {journal} {Phys. Rev. Lett.}\ }\textbf
  {\bibinfo {volume} {99}},\ \bibinfo {pages} {217405} (\bibinfo {year}
  {2007})}\BibitemShut {NoStop}%
\bibitem [{\citenamefont {Fainstein}\ \emph {et~al.}(2013)\citenamefont
  {Fainstein}, \citenamefont {Lanzillotti-Kimura}, \citenamefont {Jusserand},\
  and\ \citenamefont {Perrin}}]{Fainstein2013}%
  \BibitemOpen
  \bibfield  {author} {\bibinfo {author} {\bibfnamefont {A.}~\bibnamefont
  {Fainstein}}, \bibinfo {author} {\bibfnamefont {N.~D.}\ \bibnamefont
  {Lanzillotti-Kimura}}, \bibinfo {author} {\bibfnamefont {B.}~\bibnamefont
  {Jusserand}}, \ and\ \bibinfo {author} {\bibfnamefont {B.}~\bibnamefont
  {Perrin}},\ }\bibfield  {title} {\enquote {\bibinfo {title} {{S}trong
  {O}ptical-{M}echanical {C}oupling in a {V}ertical {G}a{A}s/{A}l{A}s
  {M}icrocavity for {S}ubterahertz {P}honons and {N}ear-{I}nfrared {L}ight},}\
  }\href {\doibase 10.1103/PhysRevLett.110.037403} {\bibfield  {journal}
  {\bibinfo  {journal} {Phys. Rev. Lett.}\ }\textbf {\bibinfo {volume} {110}},\
  \bibinfo {pages} {037403} (\bibinfo {year} {2013})}\BibitemShut {NoStop}%
\bibitem [{\citenamefont {Jusserand}\ \emph {et~al.}(2015)\citenamefont
  {Jusserand}, \citenamefont {Poddubny}, \citenamefont {Poshakinskiy},
  \citenamefont {Fainstein},\ and\ \citenamefont {Lemaitre}}]{Jusserand2015}%
  \BibitemOpen
  \bibfield  {author} {\bibinfo {author} {\bibfnamefont {B.}~\bibnamefont
  {Jusserand}}, \bibinfo {author} {\bibfnamefont {A.~N.}\ \bibnamefont
  {Poddubny}}, \bibinfo {author} {\bibfnamefont {A.~V.}\ \bibnamefont
  {Poshakinskiy}}, \bibinfo {author} {\bibfnamefont {A.}~\bibnamefont
  {Fainstein}}, \ and\ \bibinfo {author} {\bibfnamefont {A.}~\bibnamefont
  {Lemaitre}},\ }\bibfield  {title} {\enquote {\bibinfo {title} {Polariton
  resonances for ultrastrong coupling cavity optomechanics in
  $\mathrm{GaAs}/\mathrm{AlAs}$ multiple quantum wells},}\ }\href {\doibase
  10.1103/PhysRevLett.115.267402} {\bibfield  {journal} {\bibinfo  {journal}
  {Phys. Rev. Lett.}\ }\textbf {\bibinfo {volume} {115}},\ \bibinfo {pages}
  {267402} (\bibinfo {year} {2015})}\BibitemShut {NoStop}%
\bibitem [{\citenamefont {Yu}\ and\ \citenamefont {Cardona}(2010)}]{Cardona}%
  \BibitemOpen
  \bibfield  {author} {\bibinfo {author} {\bibfnamefont {P.Y.}\ \bibnamefont
  {Yu}}\ and\ \bibinfo {author} {\bibfnamefont {M.}~\bibnamefont {Cardona}},\
  }\href {http://books.google.se/books?id=5aBuKYBT\_hsC} {\emph {\bibinfo
  {title} {{F}undamentals of {S}emiconductors: {P}hysics and {M}aterials
  {P}roperties}}},\ Graduate texts in physics\ (\bibinfo  {publisher}
  {Springer},\ \bibinfo {year} {2010})\BibitemShut {NoStop}%
\bibitem [{\citenamefont {Jusserand}(2013)}]{Jusserand2013}%
  \BibitemOpen
  \bibfield  {author} {\bibinfo {author} {\bibfnamefont {B.}~\bibnamefont
  {Jusserand}},\ }\bibfield  {title} {\enquote {\bibinfo {title} {Selective
  resonant interaction between confined excitons and folded acoustic phonons in
  {GaAs/AlAs} multi-quantum wells},}\ }\href {\doibase 10.1063/1.4817647}
  {\bibfield  {journal} {\bibinfo  {journal} {Appl. Phys. Lett.}\ }\textbf
  {\bibinfo {volume} {103}},\ \bibinfo {eid} {093112} (\bibinfo {year}
  {2013}),\ 10.1063/1.4817647}\BibitemShut {NoStop}%
\bibitem [{\citenamefont {Poddubny}\ \emph {et~al.}(2014)\citenamefont
  {Poddubny}, \citenamefont {Poshakinskiy}, \citenamefont {Jusserand},\ and\
  \citenamefont {Lema\^{\i}tre}}]{Poddubny2014}%
  \BibitemOpen
  \bibfield  {author} {\bibinfo {author} {\bibfnamefont {A.~N.}\ \bibnamefont
  {Poddubny}}, \bibinfo {author} {\bibfnamefont {A.~V.}\ \bibnamefont
  {Poshakinskiy}}, \bibinfo {author} {\bibfnamefont {B.}~\bibnamefont
  {Jusserand}}, \ and\ \bibinfo {author} {\bibfnamefont {A.}~\bibnamefont
  {Lema\^{\i}tre}},\ }\bibfield  {title} {\enquote {\bibinfo {title} {Resonant
  {B}rillouin scattering of excitonic polaritons in multiple-quantum-well
  structures},}\ }\href {\doibase 10.1103/PhysRevB.89.235313} {\bibfield
  {journal} {\bibinfo  {journal} {Phys. Rev. B}\ }\textbf {\bibinfo {volume}
  {89}},\ \bibinfo {pages} {235313} (\bibinfo {year} {2014})}\BibitemShut
  {NoStop}%
\bibitem [{\citenamefont {{Keldysh}}\ and\ \citenamefont
  {{Tikhodeev}}(1986)}]{Keldysh1986}%
  \BibitemOpen
  \bibfield  {author} {\bibinfo {author} {\bibfnamefont {L.~V.}\ \bibnamefont
  {{Keldysh}}}\ and\ \bibinfo {author} {\bibfnamefont {S.~G.}\ \bibnamefont
  {{Tikhodeev}}},\ }\bibfield  {title} {\enquote {\bibinfo {title}
  {{High-intensity polariton wave near the stimulated scattering threshold}},}\
  }\href@noop {} {\bibfield  {journal} {\bibinfo  {journal} {Sov. Phys. JETP}\
  }\textbf {\bibinfo {volume} {90}},\ \bibinfo {pages} {1852--1870} (\bibinfo
  {year} {1986})}\BibitemShut {NoStop}%
\bibitem [{Note1()}]{Note1}%
  \BibitemOpen
  \bibinfo {note} {See Supplemental Material for the derivation of the
  reflection and transmission coefficients by means of the Keldysh diagram
  technique.}\BibitemShut {Stop}%
\bibitem [{\citenamefont {Trifonov}\ \emph {et~al.}(2015)\citenamefont
  {Trifonov}, \citenamefont {Korotan}, \citenamefont {Kurdyubov}, \citenamefont
  {Gerlovin}, \citenamefont {Ignatiev}, \citenamefont {Efimov}, \citenamefont
  {Eliseev}, \citenamefont {Petrov}, \citenamefont {Dolgikh}, \citenamefont
  {Ovsyankin},\ and\ \citenamefont {Kavokin}}]{Trifonov2015}%
  \BibitemOpen
  \bibfield  {author} {\bibinfo {author} {\bibfnamefont {A.~V.}\ \bibnamefont
  {Trifonov}}, \bibinfo {author} {\bibfnamefont {S.~N.}\ \bibnamefont
  {Korotan}}, \bibinfo {author} {\bibfnamefont {A.~S.}\ \bibnamefont
  {Kurdyubov}}, \bibinfo {author} {\bibfnamefont {I.~Ya.}\ \bibnamefont
  {Gerlovin}}, \bibinfo {author} {\bibfnamefont {I.~V.}\ \bibnamefont
  {Ignatiev}}, \bibinfo {author} {\bibfnamefont {Yu.~P.}\ \bibnamefont
  {Efimov}}, \bibinfo {author} {\bibfnamefont {S.~A.}\ \bibnamefont {Eliseev}},
  \bibinfo {author} {\bibfnamefont {V.~V.}\ \bibnamefont {Petrov}}, \bibinfo
  {author} {\bibfnamefont {Yu.~K.}\ \bibnamefont {Dolgikh}}, \bibinfo {author}
  {\bibfnamefont {V.~V.}\ \bibnamefont {Ovsyankin}}, \ and\ \bibinfo {author}
  {\bibfnamefont {A.~V.}\ \bibnamefont {Kavokin}},\ }\bibfield  {title}
  {\enquote {\bibinfo {title} {Nontrivial relaxation dynamics of excitons in
  high-quality {InGaAs/GaAs} quantum wells},}\ }\href {\doibase
  10.1103/PhysRevB.91.115307} {\bibfield  {journal} {\bibinfo  {journal} {Phys.
  Rev. B}\ }\textbf {\bibinfo {volume} {91}},\ \bibinfo {pages} {115307}
  (\bibinfo {year} {2015})}\BibitemShut {NoStop}%
\bibitem [{\citenamefont {Kavokin}\ \emph {et~al.}(2006)\citenamefont
  {Kavokin}, \citenamefont {Baumberg}, \citenamefont {Malpuech},\ and\
  \citenamefont {Laussy}}]{kavbamalas}%
  \BibitemOpen
  \bibfield  {author} {\bibinfo {author} {\bibfnamefont {A.}~\bibnamefont
  {Kavokin}}, \bibinfo {author} {\bibfnamefont {J.J.}\ \bibnamefont
  {Baumberg}}, \bibinfo {author} {\bibfnamefont {G.}~\bibnamefont {Malpuech}},
  \ and\ \bibinfo {author} {\bibfnamefont {F.P.}\ \bibnamefont {Laussy}},\
  }\href@noop {} {\emph {\bibinfo {title} {{M}icrocavities}}}\ (\bibinfo
  {publisher} {Clarendon Press},\ \bibinfo {address} {Oxford},\ \bibinfo {year}
  {2006})\BibitemShut {NoStop}%
\bibitem [{\citenamefont {Prineas}\ \emph {et~al.}(2000)\citenamefont
  {Prineas}, \citenamefont {Ell}, \citenamefont {Lee}, \citenamefont
  {Khitrova}, \citenamefont {Gibbs},\ and\ \citenamefont {Koch}}]{Prineas2000}%
  \BibitemOpen
  \bibfield  {author} {\bibinfo {author} {\bibfnamefont {J.~P.}\ \bibnamefont
  {Prineas}}, \bibinfo {author} {\bibfnamefont {C.}~\bibnamefont {Ell}},
  \bibinfo {author} {\bibfnamefont {E.~S.}\ \bibnamefont {Lee}}, \bibinfo
  {author} {\bibfnamefont {G.}~\bibnamefont {Khitrova}}, \bibinfo {author}
  {\bibfnamefont {H.~M.}\ \bibnamefont {Gibbs}}, \ and\ \bibinfo {author}
  {\bibfnamefont {S.~W.}\ \bibnamefont {Koch}},\ }\bibfield  {title} {\enquote
  {\bibinfo {title} {{Exciton-polariton eigenmodes in light-coupled
  ${\mathrm{In}}_{0.04}{\mathrm{Ga}}_{0.96}\mathrm{A}\mathrm{s}/\mathrm{G}\mathrm{a}\mathrm{A}\mathrm{s}$
  semiconductor multiple-quantum-well periodic structures}},}\ }\href {\doibase
  10.1103/PhysRevB.61.13863} {\bibfield  {journal} {\bibinfo  {journal} {Phys.
  Rev. B}\ }\textbf {\bibinfo {volume} {61}},\ \bibinfo {pages} {13863--13872}
  (\bibinfo {year} {2000})}\BibitemShut {NoStop}%
\bibitem [{\citenamefont {Huynh}\ \emph {et~al.}(2006)\citenamefont {Huynh},
  \citenamefont {Lanzillotti-Kimura}, \citenamefont {Jusserand}, \citenamefont
  {Perrin}, \citenamefont {Fainstein}, \citenamefont {Pascual-Winter},
  \citenamefont {Peronne},\ and\ \citenamefont {Lema\^{\i}tre}}]{Huynh2006}%
  \BibitemOpen
  \bibfield  {author} {\bibinfo {author} {\bibfnamefont {A.}~\bibnamefont
  {Huynh}}, \bibinfo {author} {\bibfnamefont {N.~D.}\ \bibnamefont
  {Lanzillotti-Kimura}}, \bibinfo {author} {\bibfnamefont {B.}~\bibnamefont
  {Jusserand}}, \bibinfo {author} {\bibfnamefont {B.}~\bibnamefont {Perrin}},
  \bibinfo {author} {\bibfnamefont {A.}~\bibnamefont {Fainstein}}, \bibinfo
  {author} {\bibfnamefont {M.~F.}\ \bibnamefont {Pascual-Winter}}, \bibinfo
  {author} {\bibfnamefont {E.}~\bibnamefont {Peronne}}, \ and\ \bibinfo
  {author} {\bibfnamefont {A.}~\bibnamefont {Lema\^{\i}tre}},\ }\bibfield
  {title} {\enquote {\bibinfo {title} {Subterahertz phonon dynamics in acoustic
  nanocavities},}\ }\href {\doibase 10.1103/PhysRevLett.97.115502} {\bibfield
  {journal} {\bibinfo  {journal} {Phys. Rev. Lett.}\ }\textbf {\bibinfo
  {volume} {97}},\ \bibinfo {pages} {115502} (\bibinfo {year}
  {2006})}\BibitemShut {NoStop}%
\bibitem [{\citenamefont {Ivchenko}(2005)}]{Ivchenko2005}%
  \BibitemOpen
  \bibfield  {author} {\bibinfo {author} {\bibfnamefont {E.~L.}\ \bibnamefont
  {Ivchenko}},\ }\href@noop {} {\emph {\bibinfo {title} {{O}ptical
  {S}pectroscopy of {S}emiconductor {N}anostructures}}}\ (\bibinfo  {publisher}
  {Alpha Science International},\ \bibinfo {address} {Harrow, UK},\ \bibinfo
  {year} {2005})\BibitemShut {NoStop}%
\bibitem [{\citenamefont {Zyablovsky}\ \emph
  {et~al.}(2014{\natexlab{b}})\citenamefont {Zyablovsky}, \citenamefont
  {Vinogradov}, \citenamefont {Dorofeenko}, \citenamefont {Pukhov},\ and\
  \citenamefont {Lisyansky}}]{Zyablovsky_2014}%
  \BibitemOpen
  \bibfield  {author} {\bibinfo {author} {\bibfnamefont {A.~A.}\ \bibnamefont
  {Zyablovsky}}, \bibinfo {author} {\bibfnamefont {A.~P.}\ \bibnamefont
  {Vinogradov}}, \bibinfo {author} {\bibfnamefont {A.~V.}\ \bibnamefont
  {Dorofeenko}}, \bibinfo {author} {\bibfnamefont {A.~A.}\ \bibnamefont
  {Pukhov}}, \ and\ \bibinfo {author} {\bibfnamefont {A.~A.}\ \bibnamefont
  {Lisyansky}},\ }\bibfield  {title} {\enquote {\bibinfo {title} {Causality and
  phase transitions in $\mathcal{PT}$-symmetric optical systems},}\ }\href
  {\doibase 10.1103/PhysRevA.89.033808} {\bibfield  {journal} {\bibinfo
  {journal} {Phys. Rev. A}\ }\textbf {\bibinfo {volume} {89}},\ \bibinfo
  {pages} {033808} (\bibinfo {year} {2014}{\natexlab{b}})}\BibitemShut
  {NoStop}%
\bibitem [{\citenamefont {Rozas}\ \emph {et~al.}(2009)\citenamefont {Rozas},
  \citenamefont {Winter}, \citenamefont {Jusserand}, \citenamefont {Fainstein},
  \citenamefont {Perrin}, \citenamefont {Semenova},\ and\ \citenamefont
  {Lema\^{\i}tre}}]{Rozas2009}%
  \BibitemOpen
  \bibfield  {author} {\bibinfo {author} {\bibfnamefont {G.}~\bibnamefont
  {Rozas}}, \bibinfo {author} {\bibfnamefont {M.~F.~Pascual}\ \bibnamefont
  {Winter}}, \bibinfo {author} {\bibfnamefont {B.}~\bibnamefont {Jusserand}},
  \bibinfo {author} {\bibfnamefont {A.}~\bibnamefont {Fainstein}}, \bibinfo
  {author} {\bibfnamefont {B.}~\bibnamefont {Perrin}}, \bibinfo {author}
  {\bibfnamefont {E.}~\bibnamefont {Semenova}}, \ and\ \bibinfo {author}
  {\bibfnamefont {A.}~\bibnamefont {Lema\^{\i}tre}},\ }\bibfield  {title}
  {\enquote {\bibinfo {title} {Lifetime of thz acoustic nanocavity modes},}\
  }\href {\doibase 10.1103/PhysRevLett.102.015502} {\bibfield  {journal}
  {\bibinfo  {journal} {Phys. Rev. Lett.}\ }\textbf {\bibinfo {volume} {102}},\
  \bibinfo {pages} {015502} (\bibinfo {year} {2009})}\BibitemShut {NoStop}%
\bibitem [{\citenamefont {Trigo}\ \emph {et~al.}(2002)\citenamefont {Trigo},
  \citenamefont {Bruchhausen}, \citenamefont {Fainstein}, \citenamefont
  {Jusserand},\ and\ \citenamefont {Thierry-Mieg}}]{Trigo2002}%
  \BibitemOpen
  \bibfield  {author} {\bibinfo {author} {\bibfnamefont {M.}~\bibnamefont
  {Trigo}}, \bibinfo {author} {\bibfnamefont {A.}~\bibnamefont {Bruchhausen}},
  \bibinfo {author} {\bibfnamefont {A.}~\bibnamefont {Fainstein}}, \bibinfo
  {author} {\bibfnamefont {B.}~\bibnamefont {Jusserand}}, \ and\ \bibinfo
  {author} {\bibfnamefont {V.}~\bibnamefont {Thierry-Mieg}},\ }\bibfield
  {title} {\enquote {\bibinfo {title} {Confinement of acoustical vibrations in
  a semiconductor planar phonon cavity},}\ }\href {\doibase
  10.1103/PhysRevLett.89.227402} {\bibfield  {journal} {\bibinfo  {journal}
  {Phys. Rev. Lett.}\ }\textbf {\bibinfo {volume} {89}},\ \bibinfo {pages}
  {227402} (\bibinfo {year} {2002})}\BibitemShut {NoStop}%
\end{thebibliography}%


\onecolumngrid

\setcounter{equation}{0}
\setcounter{figure}{0}
\setcounter{table}{0}
\makeatletter
\renewcommand{\theequation}{S\arabic{equation}}
\renewcommand{\thefigure}{S\arabic{figure}}

\vspace{.02\paperheight}

\begin{center}
\textbf{\large Supplemental Material: Derivation of the sound transmission and reflection coefficients for a pumped quantum well by means of the Keldysh diagram technique}
\end{center}

\begin{figure}[b]
  \includegraphics[width=.45\columnwidth]{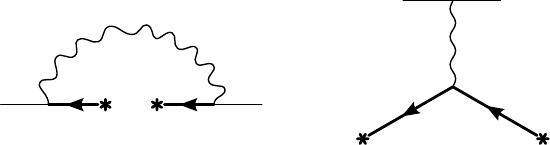}
\caption{Main contributions to the exciton self-energy $\Sigma^R$. Straight lines represent exciton, wavy lines represent phonons. An incoming (outgoing) line with star stands for laser-generated exciton polarization $b_L$ ($b_L^*$).}\label{fig:dia}
\end{figure}

First, we calculate the Green's function of a QW exciton dressed by interaction with phonons. Due to the presence of coherent pump, in addition to retarded and advanced exciton functions $G^R$ and $G^A$,  we need also Beliaev propagators $G^{\diamond}$ and $G^{\times}$, corresponding to creation of exciton pair from condensate of excitons generated by laser, and its annihilation, respectively. It is convenient to arrange these functions in a matrix~\cite{Keldysh1986}
 \begin{align}
\hat G^R (\omega)= \left[ \begin{array}{ll}
G^R(\omega)  & G^{\diamond}(\omega) \\
G^{\times}(\omega) & G^A(2\omega_L-\omega)
         \end{array} \right] \,.
\end{align}
Dyson equation for the dressed exciton Green's function reads
 \begin{align}\label{eq:sDy}
(\hat G^R)^{-1} = (\hat G^R_0)^{-1} - \hat \Sigma^R \,,
\end{align}
where bare exciton Green's function is given by
 \begin{align}
\hat G^R_0 (\omega)= \left[ \begin{array}{cc}
G^R_0(\omega)  & 0 \\
0 & G^A_0(2\omega_L-\omega)
         \end{array} \right] 
        =
\left[ \begin{array}{cc}
(\omega-\omega_x+\rmi\Gamma_x)^{-1}  & 0 \\
0 & (2\omega_L-\omega-\omega_x-\rmi\Gamma_x)^{-1}
         \end{array} \right]     
         \,,
\end{align}
while $\hat\Sigma^R$ is the exciton self-energy. The contributions of the first order in $|b_L|^2$ to the self-energy shown in Fig.~\ref{fig:dia} give
\begin{align}
\hat \Sigma^R (\omega)= \Sigma(\omega) \left[ \begin{array}{cc}
1  & \e^{2\rmi\phi} \\
\e^{-2\rmi\phi} & 1
         \end{array} \right]
         +
  \Sigma(\omega_L) \left[ \begin{array}{cc}
1  & 0 \\
0 & 1
         \end{array} \right]
\,,
\end{align}
where $\e^{\rmi\phi} = b_L/|b_L|$ and $\Sigma(\omega)$ is given by Eq.~(2) of the main text.
Solution of the Dyson equation~\eqref{eq:sDy} then gives
\begin{align}
\hat G^R (\omega)= \frac{1}{\Delta^2-(\Omega + \rmi\Gamma_x)^2 - 2\Delta\Sigma(\omega)}
  \left[ \begin{array}{cc}
\Delta-\Omega-\rmi\Gamma_x-\Sigma(\omega)  & \Sigma(\omega)\e^{2\rmi\phi}\\
\Sigma(\omega)\e^{-2\rmi\phi} & \Delta+\Omega+\rmi\Gamma_x-\Sigma(\omega)
         \end{array} \right],
\end{align}
where $\Omega = \omega - \omega_L$ and  $\Delta = \omega_L - \omega_x - \Sigma(\omega_L)$. Note that $\Sigma(\omega_L)$ is real.
 
Finally, we express transmission and reflection coefficients of a phonon with wave vector $k$ via exciton Green's functions,
\begin{align}
\left\{ \begin{array}{l} t_k-1 \\ r_k\end{array} \right\} = \left\{ \begin{array}{l}  g_k g_{k}^* \\  g_{-k} g_{k}^* \end{array} \right\} |b_L|^2 \left[  G^R(\omega_L+\Omega_k) + G^A(\omega_L-\Omega_k) + \e^{2\rmi\phi} G^\times(\omega_L+\Omega_k) +  \e^{-2\rmi\phi} G^\diamond(\omega_L+\Omega_k) \right] \,,
\end{align}
which after simplification yields
\begin{align}
&r_k = - \frac{2\rmi \Delta {\rm Im\,}\Sigma(\omega_L+\Omega_k)}{\Delta^2 - (\Omega_k+\rmi\Gamma_x)^2-2\Delta\Sigma(\omega_L+\Omega_k)}\,, \quad t_k = 1-r_k \,.
\end{align}
Neglecting the real part of $\Sigma$ while keeping ${\rm Im\,} \Sigma(\omega_L+\Omega) = - \gamma(\Omega)$ we obtain Eqs.~(5)-(6) of the main text.

%

\end{document}